\documentclass[11pt]{article}
\usepackage{graphicx,amsmath,amsfonts,amssymb,dcolumn}

\begin{document}

\title{\textbf{Aspects of nonmetricity in gravity theories}}
\author{\textbf{R.~F.~Sobreiro}\thanks{sobreiro@if.uff.br} \ , \textbf{V.~J.~Vasquez Otoya}\thanks{victor@if.uff.br}\\\\
\textit{{\small UFF $-$ Universidade Federal Fluminense,}}\\
\textit{{\small Instituto de F\'{\i}sica, Campus da Praia Vermelha,}}\\
\textit{{\small Avenida Litor\^anea s/n, 24210-346 Boa Viagem, Niter\'oi, RJ, Brasil.}}}
\date{}
\maketitle

\begin{abstract}
\noindent In this work, we show that a class of metric-affine gravities can be reduced to a Riemann-Cartan one. The reduction is based on the cancelation of the nonmetricity against the symmetric components of the spin connection. A heuristic proof, in the Einstein-Cartan formalism, is performed in the special case of diagonal unitary tangent metric tensor. The result is that the nonmetric degrees of freedom decouple from the geometry. Thus, from the point of view of isometries on the tangent manifold, the equivalence might be viewed as an isometry transition from the affine group to the Lorentz group, $A(d,\mathbb{R})\longmapsto SO(d)$. Furthermore, in this transition, depending on the form of the starting action, the nonmetricity degrees might present a dynamical matter field character, with no geometric interpretation in the Riemann-Cartan geometry.
\end{abstract}


\newpage
\section{Introduction}

The renowned article by T.~W.~B.~Kibble \cite{Kibble:1961ba} describing gravity in a gauge theoretical approach, with $ISO(d)\cong SO(d)\ltimes\mathbb{R}^d$ local symmetry group (with global translations), has inspired many consequent works on the field. In particular, the generalization of the theory to the affine group $A(d,\mathbb{R})\cong GL(d,\mathbb{R})\ltimes\mathbb{R}^d$, with global translations, has led to a new class of theories, the Metric-Affine gravities \cite{Hehl:1976kj,Hehl:1994ue}. The former \cite{Kibble:1961ba} puts gravity in a Riemann-Cartan geometry (RCG), where torsion, $T^a$, is allowed and spinorial matter can couple to gravity. The gauge fields of the theory are the vielbein\footnote{The vielbein is identified to global translations. In fact, the gauge field of translations can be associated to a part of the vielbein. See \cite{Hehl:1994ue} and references therein.}, $e^a$, associated with global translations, and the spin connection, $\omega^{ab}$, associated with the $SO(d)$ group\footnote{For mathematical purposes, always that is possible, we avoid noncompact groups in this work. Also, we call the $SO(d)$ group by Lorentz group since we are considering Euclidean tangent spaces.} and is directly related to the affine connection, $\Gamma$. The second case \cite{Hehl:1976kj,Hehl:1994ue}, describes a metric-affine geometry (MAG) where not only torsion, but also nonmetricity, $Q^{ab}$, is allowed. In this case, the relation between the spin and affine connections entails also the deviation tensor, $M^{ab}$. As it will become evident in this work, the deviation tensor is directly related to the nonmetricity.

Basically, the difference between RCG and MAG is the nonmetricity tensor, which appears as a non-vanishing quantity in the MAG. However, the nonmetricity cannot be derived from the algebra of the covariant derivatives, and thus, from the gauge procedure standpoint, it seems unnatural to consider it as a field strength. On the other hand, the nonmetricity is commonly used as a field strength or even as a propagating spin-3 physical field; see for instance \cite{Hehl:1976kj,Hehl:1994ue,Baekler:2006vw}. In that context, the nonmetricity appears explicitly in the action of a gravity theory.

Therefore, to highlight the issue of the physical role of the nonmetricity, we discuss in this work if the nonmetricity may or may not be eliminated from the geometry. To do so, we consider the full covariant derivative of the vielbein, $\mathcal{D}e=M$, where $M$ is the deviation tensor. This equation is taken here as the fundamental equation of the MAG. When $M=0$, we are dealing with the RCG \cite{Kibble:1961ba}. Further, we consider only Euclidean metrics on the tangent space. This restriction enforces a particular case of the MAG. We then show that the nonmetricity eliminates, in a natural way, the symmetric degrees of freedom of the spin connection in the full covariant derivative of the vielbein. To be more specific, the symmetric part of the deviation tensor and the symmetric part of the spin connection cancel, eliminating all the symmetric degrees of freedom of the referred equation. Thus, we have on this equation just Lorentz algebra valued quantities. The antisymmetric part of the deviation tensor is then absorbed into the antisymmetric part of the spin connection, resulting in the usual constraint for the RCG, $\widetilde{\mathcal{D}}e=0$, where $\widetilde{\mathcal{D}}$, carries just the Lorentz affine and spin connections, $\widetilde{\Gamma}$ and $\widetilde{\omega}$. Thus, as a consequence one identify a kind of geometric reduction in the form $A(d,\mathbb{R})\longmapsto SO(d)$, on the tangent manifold. Obviously, the restriction to Euclidean tangent metrics plays a fundamental hole on the cancelation of the symmetric degrees of freedom. A more general and formal proof is left for a future work \cite{Sobreiro:2009wp}.

We also discuss the consequences of this geometric transition from the curvature and torsion points of view, as well as a from a general gravity action, not explicitly depending on the nonmetricity. It is shown that the nonmetricity is not completely eliminated, but it appears as a matter field decoupled from the geometry. The same result is obtained by considering as starting point an action with an explicit dependence on the nonmetricity.

This work is organized as follows: In Sect.II, we provide a brief review of the properties, notation and conventions of the MAG, used in this work. In Sect.III, we establish a few statements in order to show the reduction of MAG to RCG. After that, the physical consequences of the reduction are discussed. Finally, in Sect.IV, we display our Conclusions.

\section{Generalities of Metric-Affine geometry}

To describe the MAG, we can think of a $d$-dimensional manifold $\mathcal{M}$ which is characterized by a metric tensor, $g$, and an affine connection, $\Gamma$, independent from each other. This is the so called metric-affine formalism. The geometry is associated with a $A(d,\mathbb{R})$ diffeomorphism invariance. We define the covariant derivative, $\nabla$, by its action on a tensor field $v$ according to
\begin{eqnarray}
\nabla_\mu v^\nu&=&\partial_\mu
v^\nu+\Gamma_{\mu\alpha}^{\phantom{\mu\alpha}\nu}v^\alpha\;,\nonumber\\
\nabla_\mu v_\nu&=&\partial_\mu
v_\nu-\Gamma_{\mu\nu}^{\phantom{\mu\nu}\alpha}v_\alpha\;,
\end{eqnarray}
where Greek indices stand for the coordinates in the manifold
$\mathcal{M}$. The curvature and torsion can be identified from
\begin{equation}
[\nabla_\mu,\nabla_\nu]v_\alpha=-R_{\mu\nu\alpha}^{\phantom{\mu\nu\alpha}\beta}v_\beta-T_{\mu\nu}^{\phantom{\mu\nu}\beta}\nabla_\beta
v_\alpha\;,
\end{equation}
where $R$ is the Riemann-Christoffel curvature and $T$ the torsion
tensor
\begin{eqnarray}
R_{\mu\nu\alpha}^{\phantom{\mu\nu\alpha}\beta}(\Gamma)&=&\partial_{[\mu}\Gamma_{\nu]\alpha}^{\phantom{\nu]\alpha}\beta}-
\Gamma_{[\mu\alpha}^{\phantom{[\mu\alpha}\gamma}\Gamma_{\nu]\gamma}^{\phantom{\nu]\alpha}\beta}\;,\nonumber\\
T_{\mu\nu}^{\phantom{\mu\nu}\alpha}&=&\Gamma_{[\mu\nu]}^{\phantom{[\mu\nu]}\alpha}\;.\label{curv1}
\end{eqnarray}
The effect of the independence between $g$ and $\Gamma$ is a
nonmetric geometry characterized by a non-trivial nonmetricity, $Q$,
\begin{equation}
Q_{\mu\nu\alpha}=\nabla_\mu g_{\nu\alpha}\;.
\end{equation}

One may also study the MAG through the isometries of the affine group in the tangent space, $\mathcal{T}$, the well-known Einstein-Cartan formalism. The gauge fields associated with translations on $\mathcal{T}$ and $GL(d,\mathbb{R})$ rotations are, respectively, the vielbein, $e$, and spin connection, $\omega$. The vielbein maps $\mathcal{M}$ quantities in $\mathcal{T}$ quantities, $v^a=e_\mu^av^\mu$. The gauge covariant derivative $D$ acts on the tangent space according to
\begin{eqnarray}
D_\mu v^a&=&\partial_\mu
v^a+\omega_{\mu\phantom{a}b}^{\phantom{\mu}a}v^b\;,\nonumber\\
D_\mu v_a&=&\partial_\mu v_a-\omega_{\mu a}^{\phantom{\mu a}b}v_b\;.\label{CD}
\end{eqnarray}
In \eqref{CD} the vielbein does not appear as a connection due to fact that, strictly speaking, it is not a connection. The vielbein is associated to the connection of translations modulo a compensating part that ensures the vector behavior of the vielbein. In fact, as discussed in \cite{Hehl:1994ue} and references therein, the equivalence between translational connections and vielbein occurs when one assumes a breaking on the local translational invariance to a global one.

From \eqref{CD} we can write
\begin{equation}
[D_\mu,D_\nu]v^a=\Omega_{\mu\nu\phantom{a}b}^{\phantom{\mu\nu}a}v^b\;,
\end{equation}
where $\Omega$ is the spin curvature
\begin{equation}
\Omega_{\mu\nu\phantom{a}b}^{\phantom{\mu\nu}a}(\omega)=\partial_{[\mu}\omega_{\nu]\phantom{a}b}^{\phantom{\nu]}a}-
\omega_{[\mu\phantom{a}c}^{\phantom{[\mu}a}\omega_{\nu]\phantom{c}b}^{\phantom{\nu]}c}\;.
\end{equation}
Further,
\begin{equation}
[D_a,D_b]v^c=\Omega_{ab\phantom{c}d}^{\phantom{ab}c}v^d-K_{ab}^{\phantom{ab}d}D_dv^c\;,
\end{equation}
where $K$ is the spin torsion
\begin{equation}
K_{ab}^{\phantom{ab}c}=e_\mu^cD_{[a}e^\mu_{b]}\;.
\end{equation}
Also, nonmetricity appears as given by
\begin{equation}
Q_{\mu}^{\phantom{\mu} ab}=D_\mu\eta^{ab}\;,\label{nm2}
\end{equation}
where $\eta$ is the flat metric tensor of $\mathcal{T}$. In general, $\eta=\eta(x)$. However, we restrict ourselves to the case that $\eta$ is strictly Euclidean. This restriction enforces a particular case of the MAG. As said before, the results on this work are restricted to this particular class of the MAG and, therefore, is taken as a heuristic proof of the geometric reduction from the MAG to the RCG.

In order to characterize the MAG by a single geometric equation, we adopt to work with the full covariant derivative, $\mathcal{D}$, acting on a $\mathcal{M}$-$\mathcal{T}$ mixed object. Here, for the sake of convenience, we take the vielbein itself,
\begin{equation}
\mathcal{D}_\mu e_\nu^a=D_\mu
e_\nu^a-\Gamma_{\mu\nu}^{\phantom{\mu\nu}\alpha}e_\alpha^a=\partial_\mu
e_\nu^a-\Gamma_{\mu\nu}^{\phantom{\mu\nu}\alpha}e_\alpha^a+\omega_{\mu\phantom{a}b}^{\phantom{\mu}a}e_\nu^b\;.\label{covder1}
\end{equation}
Notice that, with this definition,
$Q_{\mu}^{\phantom{\mu}ab}=D_\mu\eta^{ab}=\mathcal{D}_\mu\eta^{ab}$.
Now, by defining the deviation tensor $M$ as
\begin{equation}
M_{\mu\phantom{a}\nu}^{\phantom{\mu}a}=\mathcal{D}_\mu
e_\nu^a\;,\label{dev1}
\end{equation}
we rewrite expression (\ref{covder1}) as a constraint characterizing
the MAG
\begin{equation}
\partial_\mu
e_\nu^a-\Gamma_{\mu\nu}^{\phantom{\mu\nu}\alpha}e_\alpha^a+\omega_{\mu\phantom{a}b}^{\phantom{\mu}a}e_\nu^b=
M_{\mu\phantom{a}\nu}^{\phantom{\mu}a}\;,\label{covder1a}
\end{equation}
from which we can easily write the relation between the affine and spin connections
\begin{equation}
\Gamma_{\mu\nu}^{\phantom{\mu\nu}\alpha}=e^\alpha_a\partial_\mu
e_\nu^a+\omega_{\mu\phantom{a}b}^{\phantom{\mu}a}e^\alpha_ae_\nu^b-M_{\mu\phantom{\alpha}\nu}^{\phantom{\mu}\alpha}\;.\label{rel1}
\end{equation}
This constraint fixes $\Gamma$ as a function of $\omega$, $e$ and $M$. Thus, $\Gamma$ is completely determined from the properties of the tangent manifold $\mathcal{T}$ and $M$. We remark that the RCG is obtained from $M=0$. As a consequence we can interpret the deviation tensor as a measure of how the MAG differs from the RCG.

Let us develop some useful algebraic properties of the symmetry of the tangent manifold. The group decomposition of the affine group is \begin{equation}
A(d,\mathbb{R})\cong GL(d,\mathbb{R})\ltimes\mathbb{R}^d\cong
S(d)\otimes ISO(d)\;.\label{cong1}
\end{equation}
The space $S(d)$ is formally defined as the coset space,
\begin{equation}
S(d)\cong GL(d,\mathbb{R})/SO(d)\;,\label{coset1}
\end{equation}
where $S(d)$, which is not a group, can be represented by the collection of all symmetric matrices \cite{Kobayashi,Nash,Nakahara:1990th}. This space possesses $d(d+1)/2$ dimensions. The Poncar\'e group, also with $d(d+1)/2$ dimensions, is decomposed according to
\begin{equation}
ISO(d)\cong SO(d)\ltimes\mathbb{R}^d\;,\label{cong2}
\end{equation}
where $SO(d)$ is the group of pseudo-orthogonal matrices, the Lorentz group, with rank $d(d-1)/2$ and the semi-direct product with $\mathbb{R}^d$ characterizes the extra global translational symmetry.

The affine group decomposition might be used to decompose the algebra-valued spin connection. For that, we expand it on the generators of the $GL(d,\mathbb{R})$ group, $T^{ab}$,
\begin{equation}
\omega_\mu=\omega_{\mu ab}T^{ab}\;.
\end{equation}
also, $T^{ab}$ may be decomposed into the generators of the symmetric sector $\Lambda^{ab}=\Lambda^{ba}$ and the Lorentz group $\Sigma^{ab}=-\Sigma^{ba}$. Thus,
\begin{equation}
\omega_\mu=\frac{1}{2}\left(\omega_{\mu(ab)}\Lambda^{ab}+\omega_{\mu[ab]}\Sigma^{ab}\right)\;.
\end{equation}
From (\ref{nm2}) and (\ref{dev1}), we deduce that the nonmetricity is referred to
\begin{equation}
Q_{\mu ab}=\omega_{\mu(ab)}=M_{\mu(ab)}\;.\label{QW1}
\end{equation}
It is important to emphasize again that we are considering Euclidean tangent metric tensors. This condition is fundamental to obtain the first relation in \eqref{QW1}. Thus,
\begin{equation}
\omega_\mu=\frac{1}{2}\left(q_{\mu
ab}\Lambda^{ab}+\omega_{\mu[ab]}\Sigma^{ab}\right)\;,\label{decomp1}
\end{equation}
Where $q$, associated to the nonmetricity through \eqref{QW1}, is the symmetric part of the spin connection. The fact that we are restricted to Euclidean tangent metrics, implies that the vielbein transforms always through $O(d)$ group transformations. Thus, to preserve \eqref{QW1} one has to restrict the $GL(d,\mathbb{R})$ transformations to its Lorentz sector also for the symmetric part of the spin connection. Thus, a kind of symmetry breaking is enforced by the Euclidean condition on the tangent metric at the same level of the case of the vielbein and translations. This property is of remarkable importance in what follows.

\section{Reduction of MAG to RCG}

\subsection{Heuristic proof}

We now provide simple arguments concerning the relationship between the MAG and RCG. The final conclusion being that MAG and RCG are essentially equivalent to each other modulo a vector space. We start with the most general MAG based on the local affine gauge group, including local translations. Thus, we demand two \emph{ad hoc} requirements:
\begin{itemize}
\item The vielbein transforms as a vector under the Affine group transformations. As discussed at the Introduction, see \cite{Hehl:1994ue}, this requirement implies that the local translations breaks in favor of global ones.

\item The metric tensor on the tangent space are restricted to Euclidean ones, $\eta\equiv\mathbb{I}$. This second requirement is consistent with the topological nature of the $GL(d,\mathbb{R})$ group, which is noncompact. The compact sector is the $SO(d)$ subgroup. Thus, the coset space $S(d)$ is trivial and carries only trivial topological information \cite{Kobayashi,Nash,Nakahara:1990th}. Moreover, this requirement ensures the validity of \eqref{QW1} in any gauge.
\end{itemize}
Those requirements suggests that the general MAG is automatically driven to a subclass of it in which the gauge connection is the Lorentz spin connection while the vielbein and the symmetric spin connection are tensors. In fact, that is what occurs and the proof is straightforward:

\begin{itemize}
\item {\bf Statement 1}: \emph{There is no tangent manifold geometry without $SO(d)$ group}.\\
\emph{Proof}: From the decomposition (\ref{cong1}) into a trivial part and a compact one, we see that the connection character of the spin connection lives at the Lorentz sector. Also, the Lorentz group is the stability group of the affine group. This property establishes that the Lorentz group is the essential ingredient to define a geometry on the tangent manifold. Physically, it means that the Lorentz group is the sector which establishes a gauge theory for gravity. Thus, the vielbein and the symmetric part of the $GL(d,\mathbb{R})$ spin connection are taken as tensors under $SO(d)$ gauge transformations. This statement is very supportive for the two requirements above stated.

\item {\bf Statement 2}: \emph{In the full covariant derivative of the vielbein, the nonmetricity and the symmetric part of the spin connection mutually cancel}.\\
\emph{Proof}: Substituting (\ref{QW1}) in (\ref{covder1a}) we find
\begin{equation}
\partial_\mu
e_\nu^a-\Gamma_{\mu\nu}^{\phantom{\mu\nu}\alpha}e_\alpha^a+\frac{1}{2}
\left(\overline{\omega}_{\mu\phantom{a}b}^{\phantom{\mu}a}-\overline{M}_{\mu\phantom{a}b}^{\phantom{\mu}a}\right)e_\nu^b=0\;.\label{covder2}
\end{equation}
where $\overline{\omega}/2$ is the antisymmetric part of the spin
connection and $\overline{M}/2$ is the antisymmetric part of the
deviation tensor.

\item {\bf Statement 3}: \emph{The MAG can be reduced to the RCG}.\\
\emph{Proof}: The previous statement establishes that the nonmetricity and the symmetric part of the spin connection decouples from the MAG constraint (\ref{covder1a}). This means that, in (\ref{covder2}), we have just Lorentz algebra valued quantities, \emph{i.e.}, $\overline{\omega}$ and $\overline{M}$. The quantity $\widetilde{\omega}=(\overline{\omega}-\overline{M})/2$ behaves exactly as a RC spin connection, since $\overline{M}$ is a tensor. Thus, defining the RC spin connection, $\widetilde{\omega}$, according to
\begin{equation}
\widetilde{\omega}_{\mu\phantom{a}b}^{\phantom{\mu}a}=\frac{1}{2}\left(\overline{\omega}_{\mu\phantom{a}b}^{\phantom{\mu}a}-
\overline{M}_{\mu\phantom{a}b}^{\phantom{\mu}a}\right)\;,
\label{shift1}
\end{equation}
we have, from \eqref{covder2} and \eqref{shift1},
\begin{equation}
\widetilde{D}_\mu e_\nu^a=\partial_\mu
e_\nu^a-\Gamma_{\mu\nu}^{\phantom{\mu\nu}\alpha}e_\alpha^a+
\widetilde{\omega}_{\mu\phantom{a}b}^{\phantom{\mu}a}e_\nu^b=0\;,\label{covder3}
\end{equation}
which is the well-known constraint of the RCG. Thus, expression (\ref{covder1a}), characterizing the  MAG, is reduced to the expression (\ref{covder3}), characterizing the RCG. The cancelation of the nonmetric degrees of freedom with the symmetric sector of the spin connection and the redefinition of the spin connection according to (\ref{shift1}) provides then a kind of natural geometric reduction of the tangent manifold
\begin{equation}
A(d,\mathbb{R})\longmapsto SO(d)\;.\label{map1}
\end{equation}
We can then interpret the cancelation between the symmetric spin connection and the symmetric deviation tensor as an evidence of the decoupling of the nonmetric degrees of freedom from the MAG. Also, the redefinition (\ref{shift1}) seems to be compatible with background field methods \cite{'t Hooft:1975vy}, since $\overline{M}$ is Lorentz algebra valued. In expression (\ref{covder2}), the relevant quantity of the geometry is the spin connection, which is algebra-valued on the Lorentz group. The tensor field $\overline{M}$ is irrelevant for the geometry, since it can be absorbed into the spin connection. Thus, to carry $\overline{M}$ or not is just a matter of convenience. Absorbing it, we are just changing the tetrad, $e$, in other to fit it into geodesic curves. Further, in (\ref{covder2}), since there are no nonmetric degrees of freedom, one can infer that Q=0, independently of $\overline{M}$.
\end{itemize}

A more deep analysis of this section makes possible the following logic chain, more suitable for the heuristic proof: {\bf Statement 1 says that, one can reduce the affine gauge group to its compact and nontrivial sector by continuously deforming the trivial components, namely $\mathbb{R}^d$ and $S(d)$, see \cite{Kobayashi,Nash,Nakahara:1990th}. Thus, such reduction implies on the two requirements first proposed: The braking of local translations to global ones by assuming a vector behavior of the vielbein and its association with the gauge field of translations \cite{Hehl:1994ue} and also the imposition of $\eta=\mathbb{I}$, which is just the continuous deformation of the general linear group to the orthogonal subgroup. Therefore, Statements 2 and 3 follows naturally.}

\subsection{Physical discussion}

Let us now exploit the physical consequences of the geometric transition (\ref{map1}) and the spin connection redefinition (\ref{shift1}). We start with the simplest case, where the action has no explicit dependence on the nonmetricity. For that we first consider the metric-affine formalism. In this case, from (\ref{rel1}), we see that, the effect of the nonmetric degrees of freedom cancelation, together with (\ref{shift1}), on the affine connection results on the relation,
\begin{equation}
\Gamma_{\mu\nu}^{\phantom{\mu\nu}\alpha}=\widetilde{\Gamma}_{\mu\nu}^{\phantom{\mu\nu}\alpha}\;.\label{rel2}
\end{equation}
Thus, applying (\ref{rel2}) on the curvature and torsion, given in (\ref{curv1}), we find
\begin{eqnarray}
R_{\mu\nu\alpha}^{\phantom{\mu\nu\alpha}\beta}(\Gamma)&=&R_{\mu\nu\alpha}^{\phantom{\mu\nu\alpha}\beta}(\widetilde{\Gamma})\;,\nonumber\\
T_{\mu\nu}^{\phantom{\mu\nu}\alpha}(\Gamma)&=&T_{\mu\nu}^{\phantom{\mu\nu}\alpha}(\widetilde{\Gamma})\;.\label{rel4}
\end{eqnarray}
Those relations show that, if we start with a gravity theory with action $S(R,T)$, then the action is invariant under (\ref{shift1}). Thus, one can equally work in a MAG or RCG. From the metric-affine formalism point of view, both geometries are totally equivalent (when no explicit terms on nonmetricity are considered).

We can also analyze the previous effect from the Einstein-Cartan formalism. In this case, the relations (\ref{decomp1}) and (\ref{shift1}) provide
\begin{equation}
\omega=\frac{1}{2}(q+\overline{\omega})=M+\widetilde{\omega}
\end{equation}
Thus
\begin{eqnarray}
\Omega_{\mu\nu a}^{\phantom{\mu\nu a}b}(\omega)&=&\Omega_{\mu\nu
a}^{\phantom{\mu\nu a}b}(q/2+\overline{\omega}/2)\;,\nonumber\\
K_{\mu\nu}^{\phantom{\mu\nu}a}(\omega)&=&K_{\mu\nu}^{\phantom{\mu\nu}a}(q/2+\overline{\omega}/2)\;.\label{rel3}
\end{eqnarray}
In this case, things are not so easy as in the metric-affine formalism. From the tangent manifold isometries, the transition (\ref{map1}) costs the explicit appearance of the symmetric degrees of freedom of the $GL(d,\mathbb{R})$ spin connection. However, the interpretation of this effect is not difficult: {\bf There are two equivalent possibilities to work with. The first is to work in the MAG and deal with nonmetric properties of the geometry. The second is to work in a metric geometry, the RCG, with the dynamical field $q$, which has no geometric interpretation after the transition (\ref{map1}). In this case, $q$ behaves as a matter field living in a RCG}.

It remains to discuss what happens if a gravity action depends explicitly on the nonmetricity. In this case, a general gauge invariant dependence on the nonmetricity will not be eliminated by the reduction \eqref{map1}. In general, nonmetricity terms survive the transition \eqref{map1}. However, as in the previous case, no geometrical interpretation is assembled to $q$. In fact, $q$, again, is just a matter field embedded in a RCG.

\section{Conclusions}

In this article we have discussed the possibility of reducing the metric-affine geometry to a Riemann-Cartan one. This relation follows from the cancelation of the symmetric $GL(d,\mathbb{R})$ spin connection against the symmetric part of the deviation tensor, in the full covariant derivative of the vielbein. The effect is the transition described in (\ref{map1}).

On physical grounds, the transition (\ref{map1}) does not affect neither the affine nor Lorentz symmetries; it simply shows that the non-$SO$ degrees of freedom decouple from the theory, resulting on a pure Lorentz gauge theory \emph{a l\'a} Kibble \cite{Kibble:1961ba}. In terms of a gravity action with explicit nonmetricity dependence, to maintain the validity of our results, the nonmetricity should always appear in the combination $\omega-q/2$. Otherwise $q$ manifests itself as a dynamical matter field coupled to a RCG. Thus, there would be, essentially, two possibilities for constructing a gravity theory: a.) Accept (\ref{map1}) and construct theories with no $Q$ dependence at the Riemann-Cartan sector. In that case, the theory would be totally equivalent in both geometrical descriptions. b.) Start with a general nonmetric geometry and deal with a dynamical matter field after the reduction \eqref{map1}.

We emphasize that the main result of this work, described by the geometric reduction \eqref{map1} is based on the heuristic approach of Sect.~3.1. A formal analysis of the reduction \eqref{map1} and a rigorous proof, and also their physical consequences, will be discussed in a future work \cite{Sobreiro:2009wp}.

\section*{Acknowledgements}

We are indebted with A.~Accioly, J.~A.~Helay\"el-Neto and I.~D.~Soares for fruitful discussions and suggestions on this work as well as for reviewing the text. F.~Hehl and M.~M.~Sheikh-Jabbari are acknowledge for the kind correspondence. Also, the authors express their gratitude to the Conselho Nacional de Desenvolvimento Cient\'{i}fico e Tecnol\'{o}gico (CNPq-Brazil) for the financial support. Still, the authors would like to acknowledge the CBPF where this work has started in 2007. Finally, RFS would like to acknowledge the SR2-UERJ and the Theoretical Physics Department at UERJ, where part of this work was developed in 2008.


\begin{thebibliography}{99}

\bibitem{Kibble:1961ba}
  T.~W.~B.~Kibble,
  J.\ Math.\ Phys.\  {\bf 2}, 212 (1961).

\bibitem{Hehl:1976kj}
  F.~W.~Hehl, P.~Von Der Heyde, G.~D.~Kerlick and J.~M.~Nester,
  Rev.\ Mod.\ Phys.\  {\bf 48}, 393 (1976).

\bibitem{Hehl:1994ue}
  F.~W.~Hehl, J.~D.~McCrea, E.~W.~Mielke and Y.~Ne'eman,
  Phys.\ Rept.\  {\bf 258}, 1 (1995)
  [arXiv:gr-qc/9402012].

\bibitem{Baekler:2006vw}
  P.~Baekler, N.~Boulanger and F.~W.~Hehl,
  Phys.\ Rev.\  D {\bf 74}, 125009 (2006)
  [arXiv:hep-th/0608122].

\bibitem{Sobreiro:2009wp}
  R.~F.~Sobreiro and V.~J.~Vasquez Otoya,
  Work in progress.

\bibitem{Kobayashi}
  S.~Kobayashi and K.~Nomizu,
  ``Foundations of Differential Geometry, Vol. 1''
  {\it New York, USA: John Wiley \& Sons (1963).}

\bibitem{Nash}
  C.~Nash and S.~Sen,
  ``Topology and Geometry for Physicists''
  {\it London, UK: Academic Press (1983).}

\bibitem{Nakahara:1990th}
  M.~Nakahara,
  ``Geometry, topology and physics,''
{\it  Bristol, UK: Hilger (1990) 505 p. (Graduate student series in
physics).}

\bibitem{'t Hooft:1975vy}
  G.~'t Hooft,
{\it  In *Karpacz 1975, Proceedings, Acta Universitatis
Wratislaviensis No.368, Vol.1*, Wroclaw 1976, 345-369.}

\end{thebibliography}
\end{document}